\documentclass[journal]{IEEEtran}
\usepackage{amsfonts}
\usepackage{amssymb,bm,graphicx,subfigure}

\ifCLASSINFOpdf
\else
\fi
\usepackage[cmex10]{amsmath}
\begin{document}
%
\title{Frequency Estimation of Multiple Sinusoids with Sub-Nyquist Sampling Sequences}
%
%
%
\author{Shan~Huang,~
        Hong~Sun,~
        Haijian~Zhang,~
        and~Lei~Yu
\thanks{The authors are with Signal Processing Laboratory, School of Electronic Information, Wuhan University, Wuhan 430072, China (e-mail: staronice@whu.edu.cn; hongsun@whu.edu.cn; haijian.zhang@whu.edu.cn; ly.wd@whu.edu.cn).}
\thanks{}
}

%
%

\markboth{Journal of \LaTeX\ Class Files,~Vol.~11, No.~4, December~2012}%
{Shell \MakeLowercase{\textit{et al.}}: Bare Demo of IEEEtran.cls for Journals}
%



\maketitle

\begin{abstract}
In some applications of frequency estimation, the frequencies of multiple sinusoids are required to be estimated from sub-Nyquist sampling sequences. In this paper, we propose a novel method based on subspace techniques to estimate the frequencies by using under-sampled samples. We analyze the impact of under-sampling and demonstrate that three sub-Nyquist sequences are general enough to estimate the frequencies under some condition. The frequencies estimated from one sequence are unfolded in frequency domain, and then the other two sequences are used to pick the correct frequencies from all possible frequencies. Simulations illustrate the validity of the theory. Numerical results show that this method is feasible and accurate at quite low sampling rates.
\end{abstract}


%
\IEEEpeerreviewmaketitle

\section{Introduction}
%
%
%
%
\IEEEPARstart{F}{requency} estimation of multiple sinusoids has wide applications in communications, audio, medical instrumentation and electric systems \cite{luise1995carrier}\cite{kia2007high}\cite{duan2010multiple}. Frequency estimation methods cover classical modified DFT \cite{candan2015fine}, subspace techniques such as MUSIC \cite{schmidt1986multiple} and ESPRIT \cite{roy1989esprit} and other advanced spectral estimation approaches \cite{stoica1989maximum}. In general, the sampling rate of the signal is required to be higher than twice the highest frequency (i.e. Nyquist rate). The sampling frequency increases as the frequencies of the signals, which results in much hardware cost in applications \cite{hill1994benefits}. Moreover, in some applications, such as velocity synthetic aperture radar (VSAR) \cite{friedlander1998vsar}, the received signals may be of undersampled nature. So it is necessary to estimate the frequencies of the signals from undersampled measurements. In addition, this problem has a close connection with phase unwrapping in radar signal processing and sensor networks \cite{li2008fast}.

A number of methods have been proposed to estimate the frequencies with sub-Nyquist sampling. To avoid the frequency ambiguity, Zoltowski proposed a time delay method which requires the time delay difference of the two sampling channels less than or equal to the Nyquist sampling interval \cite{zoltowski1994real}. By introducing properly chosen delay lines, and by using sparse linear prediction, the method in \cite{tufts1995digital} provided unambiguous frequency estimates using low A/D conversion rates. The authors of \cite{li2009robust} made use of Chinese Remainder Theorem (CRT) to overcome the ambiguity problem, but only single frequency determination is considered. Bourdoux used the non-uniform sampling to estimate the frequency \cite{bourdoux2011sparse}. Some scholars used multi-channel sub-Nyquist sampling with different sampling rates to obtain unique signal reconstruction \cite{venkataramani2000perfect}\cite{sun2012wideband}. These methods usually have restriction on the number of the frequency components, which depends on the number of the channels. Based on emerging compressed sensing theory, sub-Nyquist wideband sensing algorithms and corresponding hardware were designed to estimate the power spectrum of a wideband signal \cite{tropp2010beyond}\cite{mishali2010theory}\cite{tian2012cyclic}. However, these methods usually require random samples, which often leads to complicated hardware, making the practicability discounted.

In this paper, we propose a new method to estimate the frequency components from under-sampled measurements. When the sampling rates satisfy some conditions, three sub-Nyquist sequences are enough to estimate the frequencies in general. The method only requires three under-sampled channels without additional processing, so the hardware is simpler than the most of former methods. The paper is organized as follows: Section II gives our analysis and method. Simulation results are shown in Section III. The last section draws conclusions.

\section{Preliminaries }
\subsection{Problem Formulation}
Consider a complex signal $x(t)$ containing $K$ frequency components with unknown constant amplitudes and phases, and additive noise that is assumed to be a zero-mean stationary complex white Gaussian random process. The samples of the signal at the sampling rate $f_S=1/\Delta t$ can be written as
\begin{equation}\label{eq1}
 x(n) =\sum\limits_{k = 1}^K {{ s_k}{e^{j(2\pi {f_k}n\Delta t)}} + w(n)},n=1,2,\cdots,
\end{equation}
where $f_k$ is the $k$-th frequency, $s_k$ is the corresponding complex amplitude, and $ w(n)$ is additive Gaussian noise. Assume that the upper limit of the frequencies is known, but we only have low-rate analog-to-digital converters whose sampling rates are far lower than the Nyquist rate. Many articles use multi-channel measurement systems to estimate the frequencies. We shall demonstrate that three-channel sub-Nyquist sampling with specific rates is enough in general.
\subsection{The Unfolding in Frequency Domain}
Suppose the highest frequency contained in the signal is lower than $f_{H}$, we sample at the rate $f_{S1}=f_{H}/a~(a>1)$. The obtained samples can be written as
\begin{equation}\label{eq2}
x(n) = \sum\limits_{k = 1}^K {{s_k}{e^{j2\pi {f_k} \cdot na/{f_H}}} + w(n)} ,n = 1,2, \cdots.
\end{equation}
As a super-resolution subspace technique, ESPRIT has favourable robustness to noise \cite{roy1989esprit}. So we choose ESPRIT as the first step of our method. The detailed process is omitted here. After using ESPRIT for one sub-Nyquist sample sequence, we obtain $K$ generalized eigenvalues $\eta_{k}(k=1,2,\cdots,K)$ corresponding to different $f_{k}$, i.e. $\eta_{k} = {e^{j2\pi {f_k} \cdot a/{f_H}}}$.\footnote{Here we assume that the generalized eigenvalues are in one-to-one correspondence with the frequencies.} Denoting the principal argument of $\eta_{k}$ as $Arg(\eta_{k})$, we obtain the estimation from the sub-Nyquist samples
\begin{equation}\label{eq3}
  {\hat{f}_k}={Arg(\eta)\cdot f_{H}}/{(2\pi a)}, k=1,2,\cdots,K.
\end{equation}
In the case of under-sampling the values of $a\cdot f_k/f_H$ may be greater than 1 for some $k$. Due to the periodicity of trigonometric functions, a series of estimated frequencies $\tilde{f}_{k}$ all satisfy ${e^{j2\pi {\tilde{f}_k} \cdot a/{f_H}}}=\eta_{k}$. The $\hat{f}_k$ are only a part of possible frequencies in the bottom of the frequency range $(0,f_H)$. We restrict the under-sampled ratio $a$ to be a positive integer. All the possible frequencies (we call eligible frequencies) can be unfolded as
\begin{equation}\label{eq4}
  {\tilde f_k} = {{{\hat f}_k} + \alpha_{k} \cdot {f_H}}/a,~\alpha_{k} = 0,1, \cdots ,a - 1.
\end{equation}
\subsection{The Matching of The Frequencies}
Obviously, it's almost impossible to determine the correct frequencies through only one sub-Nyquist sequence. If another sequence sampled at the rate $f_{S2}=f_{H}/b~(b>1)$ is utilized, another set of eligible frequencies can be obtained, namely
\begin{equation}\label{eq5}
  {\tilde f'_k} = {{{\hat f'}_k} + \beta_{k} \cdot {f_H}}/b,~\beta_{k} = 0,1, \cdots ,b - 1,
\end{equation}
where $(\ast)'$ denotes the parameters related to the second sample sequence. The sets of the eligible frequencies ${\tilde f_k}$ and ${\tilde f'_k}$ contain the same group, which is composed of the correct frequencies. If ${\tilde f_k}$ and ${\tilde f'_k}$ can be matched one to one, the correct frequencies will be found. Let ${\tilde f_m}={\tilde f'_l}$, we have
\begin{equation}\label{eq6}
   {{\hat f}_m} + {\alpha_m} \cdot {f_H} /a = {{\hat f'}_l}  + {{\beta}_l} \cdot {f_H} /b,
\end{equation}
i.e.,
\begin{equation}\label{eq7}
  b{\alpha_m} - a{\beta_l} = ab\left( {{{\hat f'}_l} - {{\hat f}_m}} \right)/{f_H}.
\end{equation}
According to B\'{e}zout's identity, when $a$ is coprime to $b$ (i.e. $a\perp b$), (\ref{eq7}) has integer solutions $\alpha_{m},\beta_{l}$ as long as its right hand side is an integer. Moreover, since $\alpha_m\in \{0,1,\cdots,a-1\}$ and ${\beta}_{l}\in \{0,1,\cdots,b-1\}$, the equation (\ref{eq7}) just has an unique satisfactory solution. Denoting the true values of $\alpha_{m},{\beta_l}$ by $\bar{\alpha}_{m},\bar{\beta}_l$, we have
\begin{equation}\label{eq8}
   f_{m}={{{\hat f}_m} + \bar{\alpha}_{m} \cdot {f_H}}/a,~f_{l}={{{\hat f'}_l} + \bar{\beta}_{l} \cdot {f_H}}/b.
\end{equation}
So the right hand side of (\ref{eq7}) is
\begin{equation}\label{eq9}
\begin{split}
  ab\left( {{{\hat f'}_l} - {{\hat f}_m}} \right)/{f_H} =ab\left( {{f_l} - {f_m}} \right)/{f_H} + b{{\bar \alpha}_m} - a{{\bar \beta}_l}.
\end{split}
\end{equation}
To make the matching succeed, (\ref{eq9}) should be an integer when and only when $l=m$. It can be seen that when $\Delta f\triangleq f_l-f_m(l\neq m)$ is an integer multiple of $f_H/\left(ab\right)$, the right hand side of (\ref{eq7}) must be an integer. That is to say, when two true frequencies $f_l,f_m$ satisfy that $\Delta f\triangleq f_l-f_m(l\neq m)$ is an integer multiple of $f_H/\left(ab\right)$, the matching of the eligible frequencies may suffer from mistakes.

Now suppose that we have three sub-Nyqiust sequences which are sampled at $f_H/a,f_H/b$ and $f_H/c$ respectively and $a,b,c$ are pairwise coprime. When any two sequences are used to estimate the frequencies of the signal, the matching of the eligible frequencies may fail when there exist $f_l-f_m=pf_H/ab$ or $f_l-f_m=qf_H/ac$ or $f_l-f_m=rf_H/bc~(p,q,r\in \mathbb{Z})$. Consider that the three sub-Nyquist sequences are jointly used to estimate the frequencies. Since $0\leq f_k< f_H$, there exist no such $f_l,f_m(l\neq m)$ that satisfy $f_l-f_m=pf_H/ab$ and $f_l-f_m=qf_H/ac$ and $f_l-f_m=rf_H/bc$ simultaneously. So the results obtained from arbitrary two sequences can complement each other, making the probability of the mismatch much lower. That is to say, three sub-Nyquist sample sequences are enough to guarantee the success of the matching in general. According to our numerical results, few values of the true frequencies lead to wrong results.
\begin{figure*}[!htbp]
\centering
\includegraphics[scale=0.9]{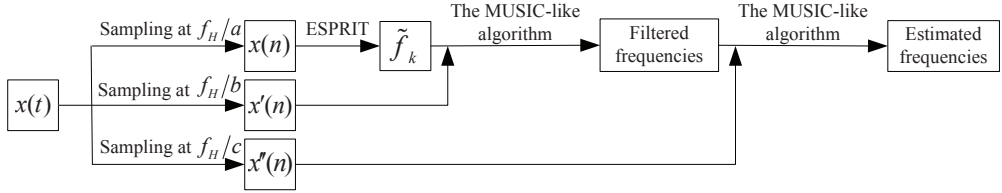}
\centering
\caption{ The procedure diagram of the method. } \label{fig1}
\end{figure*}
\subsection{The Algorithm for Screening}
After the first step, the estimated frequencies via ESPRIT for one sub-Nyquist sequence are unfolded in frequency domain and all possible frequencies are obtained. Instead of the complicated matching process, we use a MUSIC-like algorithm to screen the correct frequencies from the eligible frequencies. The algorithm for screening utilizes subspace techniques and plays the same role with the matching process. Conventional MUSIC uses fixed uniform frequency grids to form the steering vectors, while we use the eligible frequencies to form the steering vectors in the algorithm.

The second sample sequence can be written as
\begin{equation}\label{eq10}
  x'(n)= \sum\limits_{k = 1}^K {{s_k}{e^{j2\pi bn \left({{\hat f}_k}/{f_H} + {{\bar \alpha}_k} /a\right)}}}+ w'(n).
\end{equation}
Let $g_k={\hat f}_k/f_H$, the time series of the second sampling is expressed as
\begin{equation}\label{eq11}
  \bm x' = \left[\bm v_{1},\bm v_{2}, \cdots ,\bm v_{K}\right]\bm s +\bm w',
\end{equation}
where
\begin{align}\label{eq12}
\bm x' &=[x(n),\cdots,x(n+N-1)]^{T},\\
\bm w' &=[w(n),\cdots,w(n+N-1)]^{T},\\
\bm s &= {\left[ {{s_1}{e^{j2\pi {f_1}n\Delta t}}, \cdots ,{s_K}{e^{j2\pi {f_K}n\Delta t}}} \right]^T},\\
\bm v_k &= \left[ e^{j2\pi b( g_k +{\bar \alpha}_k/a )}, \cdots ,e^{j2\pi bN( g_k +{\bar \alpha}_k/a )} \right]^T,
\end{align}
and $N$ is the number of the sequential samples, $[\ast]^{T}$ denotes the transpose operation. For each $k$, $\bm v_{k}$ is augmented as
\begin{equation}\label{eq13}
\bm v(g_k) = {\left[ {{\bm v}({g_k},0),{\bm v}({g_k},1), \cdots ,{\bm v}({g_k},a-1)} \right]},
\end{equation}
where
\begin{equation}\label{eq14}
  \bm {v}{(g_k,\alpha_{k})} = {\left[ {e^{j2\pi b( g_k + \alpha_k/a)}}, \cdots , {e^{j2\pi bN( g_k + \alpha_k/a)}}\right]^T}.
\end{equation}
So (\ref{eq11}) changes into
\begin{equation}\label{eq15}
  \bm x' = \bm V\bm {\tilde{s}} + \bm w',
\end{equation}
where
\begin{multline}\label{eq16}
 \bm V =  [{\bm v}({g_1},0),{\bm v}({g_1},1) \cdots ,{\bm v}({g_1},a-1),\cdots ,\\{\bm v}({g_K},0),{\bm v}({g_K},1) \cdots ,{\bm v}({g_K},a-1)] ,
\end{multline}
and
\begin{multline}\label{eq17}
  \bm {\tilde{s}} = [ {s_1}{e^{j2\pi {f_1}n\Delta t}}, \cdots ,{s_1}{e^{j2\pi {f_1}n\Delta t}}, \cdots ,\\{s_K}{e^{j2\pi {f_K}n\Delta t}}, \cdots ,{s_K}{e^{j2\pi {f_K}n\Delta t}}] ^T.
\end{multline}

The following procedure is similar to MUSIC. Taking the eigen-decomposition of the auto-covariance matrix of $\bm x'$ renders
\begin{equation}\label{eq18}
  {\bm R_{{\bm{x'x'}}}} = \left[ {{\bm U_s},{\bm U_e}} \right]\left[ {\begin{array}{*{20}{c}}
\bm \Lambda &0\\
0&{{{\sigma '}^2}{\bm I}}
\end{array}} \right]\left[ {\begin{array}{*{20}{c}}
{\bm U_s^H}\\
{\bm U_e^H}
\end{array}} \right],
\end{equation}
where the column vectors of $\bm U_{s}$ and $\bm U_e$ are, respectively, the eigenvectors that span the signal subspace and the noise subspace of $\bm R_{\bm{x'x'}}$ with the associated eigenvalues on the diagonals of $\bm \Lambda$ and ${\sigma '}^2{\bm I}$ ($\bm I$ is the identity matrix), and $[\ast]^{H}$ denotes the conjugate transpose operation. So the pseudo-spectrum with respect to $\tilde{f}_{k}= g_k\cdot f_H+\alpha_k \cdot f_H/a$ can be given by
\begin{equation}\label{eq19}
{P_{MU}}({\tilde f_k}) = \frac{1}{{{\bm v}({g_k},\alpha_{k}){\bm U_e}\bm U_e^H{\bm v}({g_k},\alpha_{k})}},
\end{equation}
where $k=1,2,\cdots,K$, $\alpha_{k} =0,1,\cdots,a-1$. The maximum several wave peaks indicate the correct frequencies. In (\ref{eq14}), if $N$ is an integral multiple of $a$, the steering vectors ${\bm v}({g_k},\alpha_{k})$ with respect to the same ${\hat f}_k$ are mutually orthogonal, which makes it easy to distinguish the eligible frequencies. So the condition that $a$ and $b$ are required to be relatively prime also provides high discriminability for screening.

We have demonstrated that three sub-Nyquist sequences with specific sampling rates are general enough to determine the correct frequencies. The three sub-Nyquist sequences are all necessary to guarantee the success of the screening. We should use the MUSIC-like algorithm for the other two sequences respectively and find the intersecting frequencies. Or another approach is filtering out the false frequencies twice via the MUSIC-like algorithm and the correct frequencies are remained. The block diagram of the latter approach is shown in Fig.~\ref{fig1}. It is worth mentioning that different true frequencies may correspond to the same ${\hat{f}_k}$ in the first step. This situation is beyond the above discussion. It may disturb the matching procedure but not necessarily ruin the algorithms. In simulations, we often find that this situation has no influence on the final result because no correct frequencies are left out from the eligible frequencies. Moreover, if the frequencies distribute randomly in some frequency range, the probability of this situation is quite low.
\section{Simulation Results }
Firstly we make experiments to verify the conclusion that three sub-Nyquist sequences are enough to estimate the frequencies. For the sake of brevity, $a=3,b=4,c=5$ and $f_H=60Hz$ are set, and the signals are noiseless. The signals contain $K$ frequency components with uniform amplitudes and phase angles. We call the three sequences sampled at $f_H/a,f_H/b,f_H/c$ as S1,S2,S3 respectively for convenience.

\begin{figure}[!htbp]
\centering
\subfigure[]{
\includegraphics[scale=0.6]{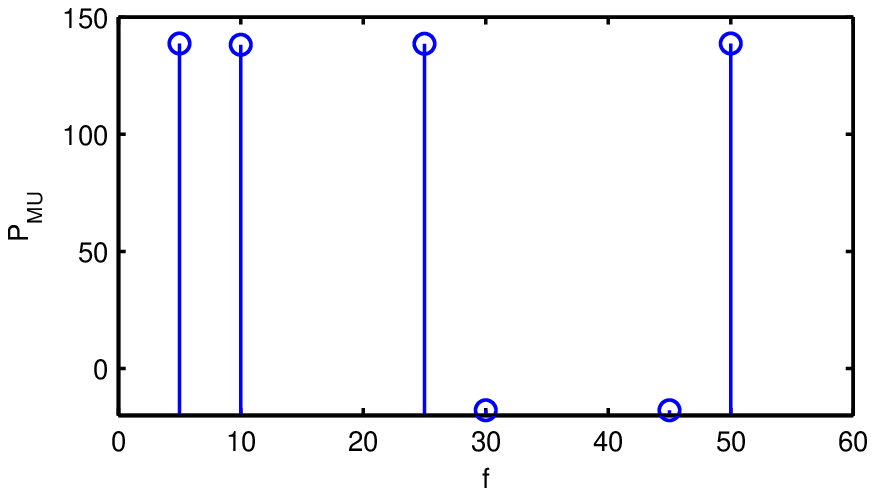}}\\
\subfigure[]{
\includegraphics[scale=0.6]{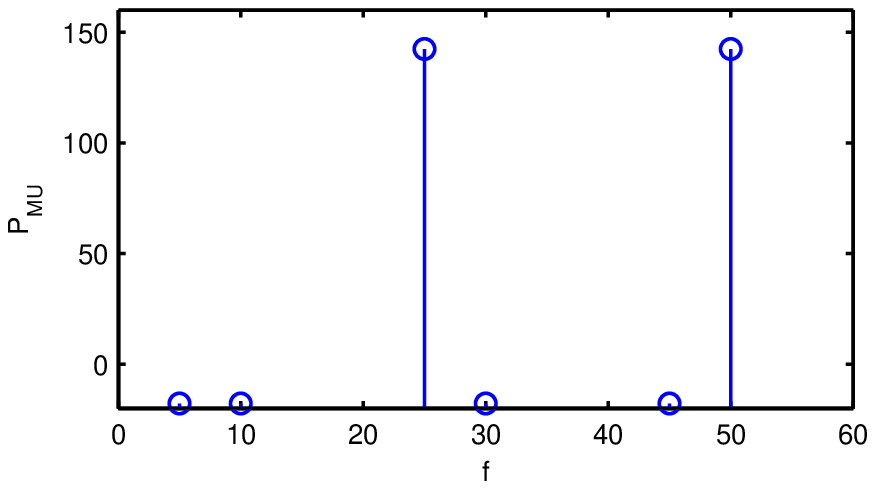}}\\
\subfigure[]{
\includegraphics[scale=0.6]{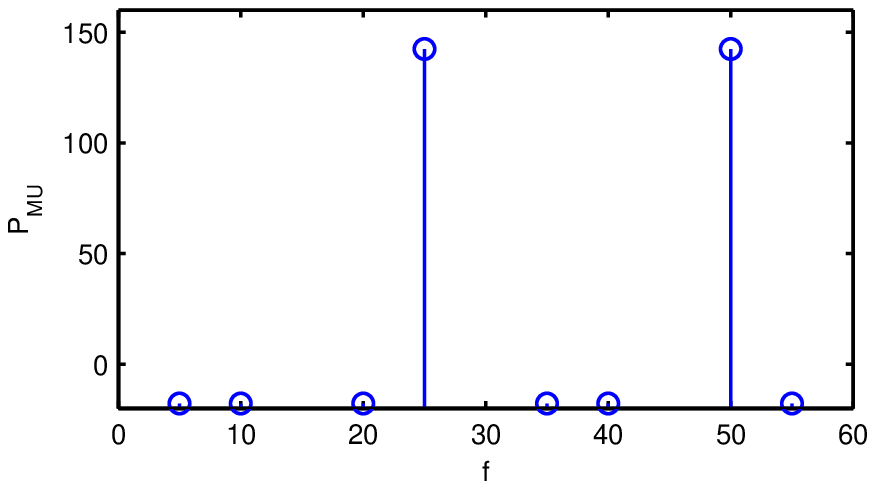}}
\caption{The pseudo-spectrum of the eligible frequencies. (a) S1,S2; (b) S1,S3; (c) S2,S3. } \label{fig2}
\end{figure}

In the first simulation, we set $K=2$ and $f_1=25Hz,f_2=50Hz$. If we use ESPRIT for S1, we will get a series of eligible frequencies, namely $5Hz, 25Hz, 45Hz, 10Hz, 30Hz, 50Hz$. Any two sequences are selected to obtain the eligible frequencies from one sequence via ESPRIT and the pseudo-spectrum of the eligible frequencies from the other sequence via the MUSIC-like algorithm. The results obtained from S1,S2 and S1,S3 and S2,S3 are all shown in Fig.~\ref{fig2}. Because $f_2-f_1=5\cdot f_H/ab$, the result obtained from S1,S2 is not able to recognize the correct frequencies as shown in Fig.~\ref{fig2}(a). While in Fig.~\ref{fig2}(b) and Fig.~\ref{fig2}(c), the correct frequencies are found. The results accord with our previous analysis.

\begin{figure}[!htbp]
\centering
\subfigure[]{
\includegraphics[scale=0.6]{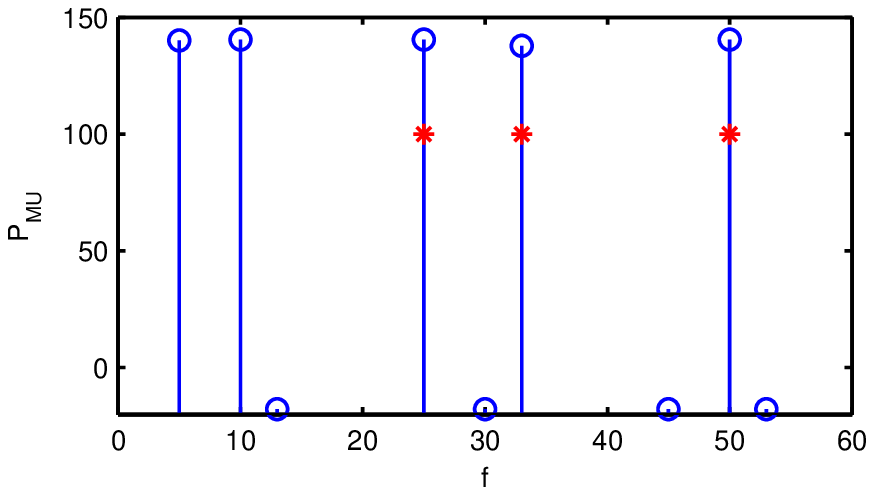}}\\
\subfigure[]{
\includegraphics[scale=0.6]{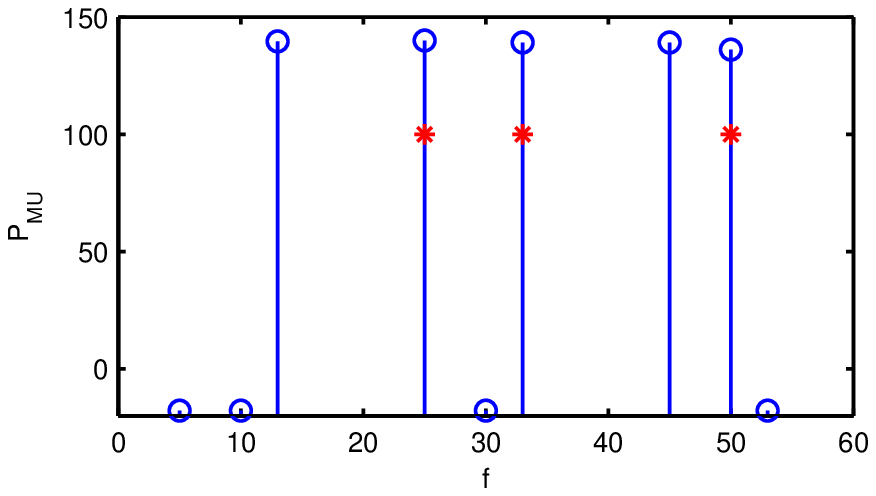}}\\
\caption{The pseudo-spectrum in two cases. (a) S1,S2; (b) S1,S3. } \label{fig3}
\end{figure}

In the second simulation, we set $K=3$ and $f_1=25Hz,f_2=33Hz,f_3=50Hz$. The results obtained from S1,S2 and S1,S3 are shown in Fig.~\ref{fig3}(a) and Fig.~\ref{fig3}(b) respectively. Although neither of the results is right, their intersection indicates the correct frequencies. This experiment validates our method in the previous section.

\begin{figure}[htbp]
\centering
\includegraphics[scale=0.7]{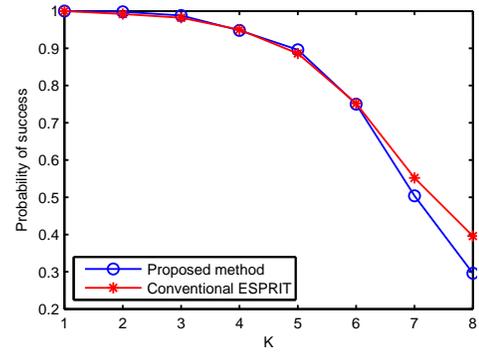}
\centering
\caption{ The probabilities of success for different numbers of frequency components. } \label{fig4}
\end{figure}

Then we test the capacity of the algorithms when the frequencies distribute uniformly in $(0,100Hz)$. The number of the frequencies $K$ varies from $1$ to $8$. The minimum interval between the frequencies is larger than $0.1Hz$ and the corresponding amplitudes are in $[0.1,1]$. $a=7,b=8,c=9$ are fixed and $100$ snapshots are used for each sub-Nyquist sequence. We do $500$ trials for each $K$ in the noisy environment with SNR=20dB. The mean square errors of estimated frequencies $f_{es}$ different from true frequencies are computed by $MSE = {{\sqrt {\sum\limits_{k = 1}^K {{{\left( {{{f}_{es}} - {f_k}} \right)}^2}} } } \mathord{\left/{\vphantom {{\sqrt {\sum\limits_{k = 1}^K {{{\left( {{{\hat f}_k} - {f_k}} \right)}^2}} } } K}} \right.\kern-\nulldelimiterspace} K}$. In this simulation, if the MSE of one trial is smaller than 0.05, we say that this is a successful trial. For the sake of comparison, we also sample the signals at the rate $f_S=f_H$ and use conventional ESPRIT to estimate the frequencies. The success probabilities of the proposed method and conventional ESPRIT for different $K$ are shown in Fig.~\ref{fig4}. The probabilities of success are higher than $90\%$ when $K\leq 5$ and roughly equivalent to conventional ESPRIT.

\begin{figure}[htbp]
\centering
\includegraphics[scale=0.7]{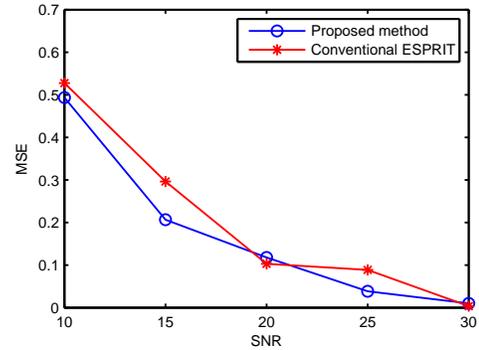}
\centering
\caption{ The average MSE for different SNR values. } \label{fig5}
\end{figure}

Finally, we compare the errors of the frequencies with conventional ESPRIT at different SNR values. $K=3$ is fixed and SNR varies from $10dB$ to $30dB$. Average values are obtained from 500 trials for each SNR. As shown in Fig.~\ref{fig5}, the proposed method has the same level of accuracy with conventional ESPRIT. This experiment demonstrates that the proposed method can achieve enough accuracy.

\section{Conclusion}
In this letter, we proposed an undersampled method based on subspace techniques to estimate the frequencies of multiple sinusoids. Three sub-Nyquist sequences at specific sampling rates are shown to be general enough for the estimation from theoretical and experimental analysis. Through the numerical simulations, we verify that the method has enough robustness and accuracy.


%


\ifCLASSOPTIONcaptionsoff
  \newpage
\fi



%
%
%

\bibliographystyle{IEEEtran}
\bibliography{IEEEabrv,mybibfile}

\begin{thebibliography}{10}
\providecommand{\url}[1]{#1}
\csname url@samestyle\endcsname
\providecommand{\newblock}{\relax}
\providecommand{\bibinfo}[2]{#2}
\providecommand{\BIBentrySTDinterwordspacing}{\spaceskip=0pt\relax}
\providecommand{\BIBentryALTinterwordstretchfactor}{4}
\providecommand{\BIBentryALTinterwordspacing}{\spaceskip=\fontdimen2\font plus
\BIBentryALTinterwordstretchfactor\fontdimen3\font minus
  \fontdimen4\font\relax}
\providecommand{\BIBforeignlanguage}[2]{{%
\expandafter\ifx\csname l@#1\endcsname\relax
\typeout{** WARNING: IEEEtran.bst: No hyphenation pattern has been}%
\typeout{** loaded for the language `#1'. Using the pattern for}%
\typeout{** the default language instead.}%
\else
\language=\csname l@#1\endcsname
\fi
#2}}
\providecommand{\BIBdecl}{\relax}
\BIBdecl

\bibitem{luise1995carrier}
M.~Luise and R.~Reggiannini, ``Carrier frequency recovery in all-digital modems
  for burst-mode transmissions,'' \emph{IEEE Transactions on Communications},
  vol.~43, no. 2/3/4, pp. 1169--1178, 1995.

\bibitem{kia2007high}
S.~H. Kia, H.~Henao, and G.-A. Capolino, ``A high-resolution frequency
  estimation method for three-phase induction machine fault detection,''
  \emph{IEEE Transactions on Industrial Electronics}, vol.~54, no.~4, pp.
  2305--2314, 2007.

\bibitem{duan2010multiple}
Z.~Duan, B.~Pardo, and C.~Zhang, ``Multiple fundamental frequency estimation by
  modeling spectral peaks and non-peak regions,'' \emph{IEEE Transactions on
  Audio, Speech, and Language Processing}, vol.~18, no.~8, pp. 2121--2133,
  2010.

\bibitem{candan2015fine}
{\c{C}}.~Candan, ``Fine resolution frequency estimation from three \text{DFT}
  samples: Case of windowed data,'' \emph{Signal Processing}, vol. 114, pp.
  245--250, 2015.

\bibitem{schmidt1986multiple}
R.~O. Schmidt, ``Multiple emitter location and signal parameter estimation,''
  \emph{IEEE Transactions on Antennas and Propagation}, vol.~34, no.~3, pp.
  276--280, 1986.

\bibitem{roy1989esprit}
R.~Roy and T.~Kailath, ``\text{ESPRIT}-estimation of signal parameters via
  rotational invariance techniques,'' \emph{IEEE Transactions on Acoustics,
  Speech and Signal Processing}, vol.~37, no.~7, pp. 984--995, 1989.

\bibitem{stoica1989maximum}
P.~Stoica, R.~L. Moses, B.~Friedlander, and T.~Soderstrom, ``Maximum likelihood
  estimation of the parameters of multiple sinusoids from noisy measurements,''
  \emph{IEEE Transactions on Acoustics, Speech and Signal Processing}, vol.~37,
  no.~3, pp. 378--392, 1989.

\bibitem{hill1994benefits}
G.~Hill, ``The benefits of undersampling,'' \emph{Electronic Design}, vol.~42,
  no.~14, p.~69, 1994.

\bibitem{friedlander1998vsar}
B.~Friedlander and B.~Porat, ``Vsar: A high resolution radar system for ocean
  imaging,'' \emph{IEEE Transactions on Aerospace and Electronic Systems,},
  vol.~34, no.~3, pp. 755--776, 1998.

\bibitem{li2008fast}
X.~Li and X.-G. Xia, ``A fast robust chinese remainder theorem based phase
  unwrapping algorithm,'' \emph{IEEE Signal Processing Letters}, no.~15, pp.
  665--668, 2008.

\bibitem{zoltowski1994real}
M.~D. Zoltowski and C.~P. Mathews, ``Real-time frequency and 2-\text{D} angle
  estimation with sub-\text{N}yquist spatio-temporal sampling,'' \emph{IEEE
  Transactions on Signal Processing}, vol.~42, no.~10, pp. 2781--2794, 1994.

\bibitem{tufts1995digital}
D.~Tufts and H.~Ge, ``Digital estimation of frequencies of sinusoids from
  wide-band under-sampled data,'' in \emph{IEEE International Conference on
  Acoustics, Speech, and Signal Processing}, vol.~5.\hskip 1em plus 0.5em minus
  0.4em\relax IEEE, 1995, pp. 3155--3158.

\bibitem{li2009robust}
X.~Li, H.~Liang, and X.-G. Xia, ``A robust \text{C}hinese remainder theorem
  with its applications in frequency estimation from undersampled waveforms,''
  \emph{IEEE Transactions on Signal Processing}, vol.~57, no.~11, pp.
  4314--4322, 2009.

\bibitem{bourdoux2011sparse}
A.~Bourdoux, S.~Pollin, A.~Dejonghe, and L.~Van~der Perre, ``Sparse signal
  sensing with non-uniform undersampling and frequency excision,'' in
  \emph{2011 Sixth International ICST Conference on Cognitive Radio Oriented
  Wireless Networks and Communications (CROWNCOM)}.\hskip 1em plus 0.5em minus
  0.4em\relax IEEE, 2011, pp. 246--250.

\bibitem{venkataramani2000perfect}
R.~Venkataramani and Y.~Bresler, ``Perfect reconstruction formulas and bounds
  on aliasing error in sub-\text{N}yquist nonuniform sampling of multiband
  signals,'' \emph{IEEE Transactions on Information Theory}, vol.~46, no.~6,
  pp. 2173--2183, 2000.

\bibitem{sun2012wideband}
H.~Sun, W.-Y. Chiu, J.~Jiang, A.~Nallanathan, and H.~V. Poor, ``Wideband
  spectrum sensing with sub-\text{N}yquist sampling in cognitive radios,''
  \emph{IEEE Transactions on Signal Processing}, vol.~60, no.~11, pp.
  6068--6073, 2012.

\bibitem{tropp2010beyond}
J.~A. Tropp, J.~N. Laska, M.~F. Duarte, J.~K. Romberg, and R.~G. Baraniuk,
  ``Beyond \text{N}yquist: Efficient sampling of sparse bandlimited signals,''
  \emph{IEEE Transactions on Information Theory}, vol.~56, no.~1, pp. 520--544,
  2010.

\bibitem{mishali2010theory}
M.~Mishali and Y.~C. Eldar, ``From theory to practice: Sub-\text{N}yquist
  sampling of sparse wideband analog signals,'' \emph{IEEE Journal of Selected
  Topics in Signal Processing}, vol.~4, no.~2, pp. 375--391, 2010.

\bibitem{tian2012cyclic}
Z.~Tian, Y.~Tafesse, and B.~M. Sadler, ``Cyclic feature detection with
  sub-\text{N}yquist sampling for wideband spectrum sensing,'' \emph{IEEE
  Journal of Selected topics in signal processing}, vol.~6, no.~1, pp. 58--69,
  2012.

\end{thebibliography}
%




\end{document}